\documentclass[manuscript,nonacm]{acmart}
\setboolean{@twoside}{false}

\usepackage{caption}
\usepackage{subcaption}
\usepackage{float}

\AtBeginDocument{%
  \providecommand\BibTeX{{%
    \normalfont B\kern-0.5em{\scshape i\kern-0.25em b}\kern-0.8em\TeX}}}

\setcopyright{acmcopyright}
\copyrightyear{2023}
\acmYear{2023}

\acmConference[JCDL '23]{ACM/IEEE Joint Conference on Digital Libraries}{June 26--30, 2023}{Santa Fe, NM}
\begin{document}

\title[Making Changes in Webpages Discoverable]{Making Changes in Webpages Discoverable: A Change-Text Search Interface for Web Archives}

\author{Lesley Frew}
\affiliation{%
  \department{\emph{Department of Computer Science}}
  \institution{\emph{Old Dominion University}}
  \city{Norfolk}
  \state{Virginia}
  \country{USA}
}
\email{lfrew001@odu.edu}
\orcid{https://orcid.org/0000-0003-0929-049X}

\author{Michael L. Nelson}
\affiliation{%
  \department{\emph{Department of Computer Science }}
  \institution{\emph{Old Dominion University}}
  \city{Norfolk}
  \state{Virginia}
  \country{USA}
}
\email{mln@cs.odu.edu}
\orcid{https://orcid.org/0000-0003-3749-8116}

\author{Michele C. Weigle}
\affiliation{%
  \department{\emph{Department of Computer Science }}
  \institution{\emph{Old Dominion University}}
  \city{Norfolk}
  \state{Virginia}
  \country{USA}
}
\email{mweigle@cs.odu.edu}
\orcid{https://orcid.org/0000-0002-2787-7166}

\renewcommand{\shortauthors}{Frew et al.}

\begin{abstract}
  Webpages change over time, and web archives hold copies of historical versions of webpages. Users of web archives, such as journalists, want to find and view changes on webpages over time. However, the current search interfaces for web archives do not support this task. For the web archives that include a full-text search feature, multiple versions of the same webpage that match the search query are shown individually without enumerating changes, or are grouped together in a way that hides changes. 
  We present a change text search engine that allows users to find changes in webpages. We describe the implementation of the search engine backend and frontend, including a tool that allows users to view the changes between two webpage versions in context as an animation. We evaluate the search engine with U.S. federal environmental webpages that changed between 2016 and 2020. The change text search results page can clearly show when terms and phrases were added or removed from webpages. The inverted index can also be queried to identify salient and frequently deleted terms in a corpus.
\end{abstract}





\maketitle

\section{Introduction}

The current ways in which we interact with the Web, including our browsers and popular search engines, predispose us to only see the most recent versions of webpages. Some webpages update their content frequently, while others remain relatively static.  Web archives preserve these historical versions of webpages. Different web archives have different holdings, and aggregating the holdings increases the number of webpage versions available for viewing. 

\begin{figure}[b]
     \centering
     \begin{subfigure}[b]{0.8\textwidth}
         \centering
         \includegraphics[width=\textwidth]{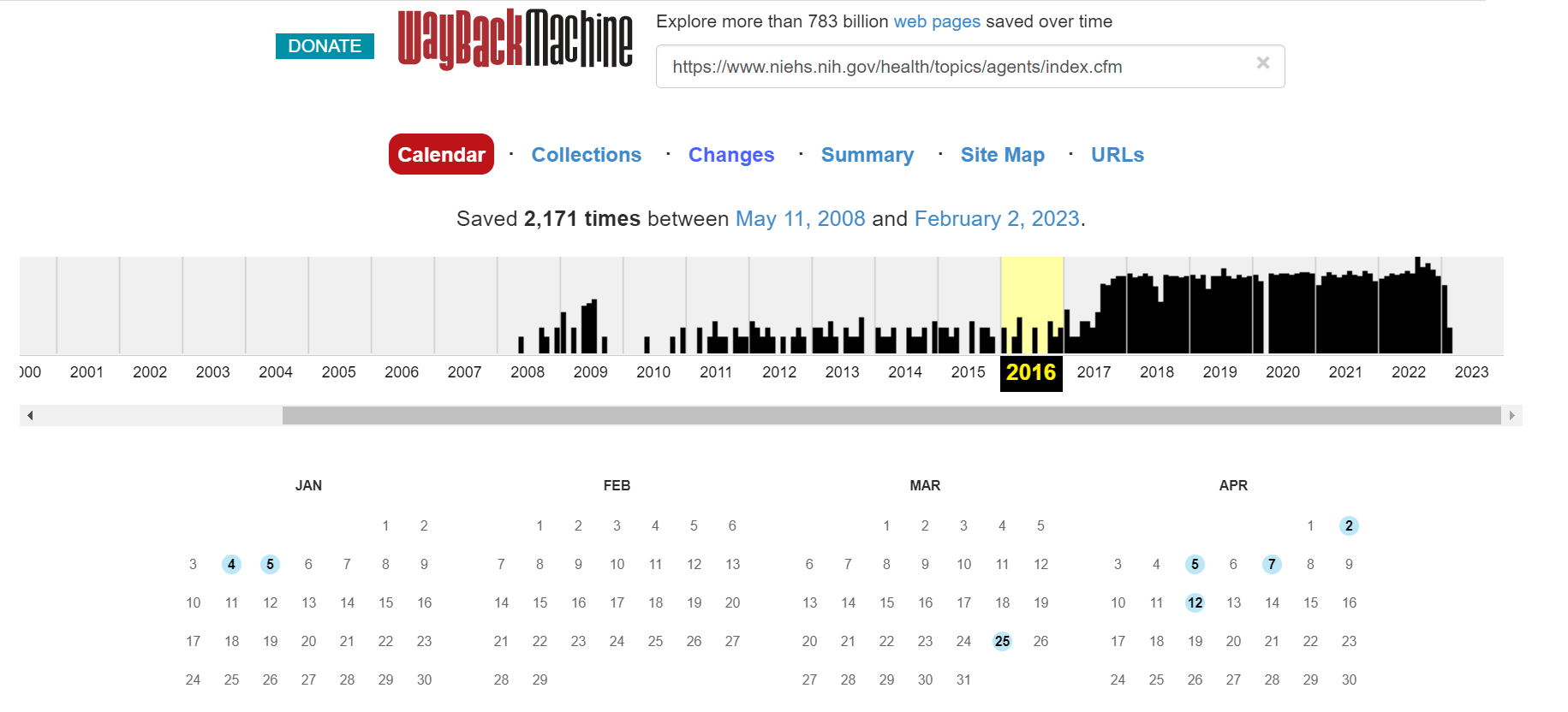}
         \caption{Internet Archive Wayback Machine URI lookup }
         \label{fig:nih1}
     \end{subfigure}
     
     \begin{subfigure}[b]{0.8\textwidth}
         \centering
         \includegraphics[width=\textwidth]{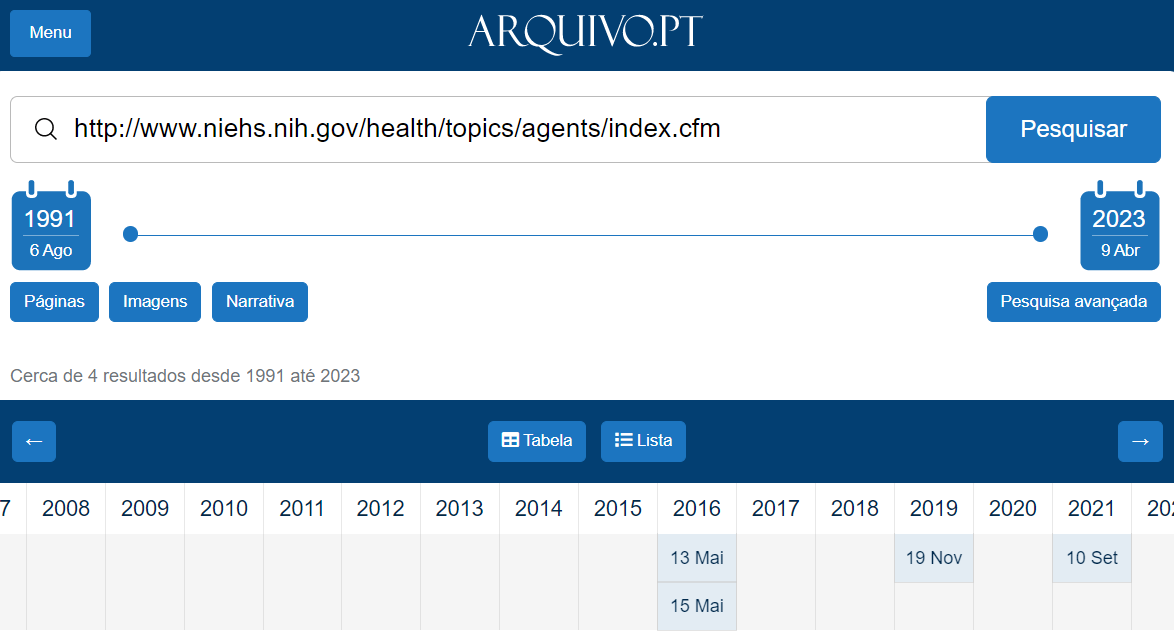}
         \caption{Portuguese Web Archive URI lookup   }
         \label{fig:nih3b}
     \end{subfigure}
        \caption{Two examples of web archive URI lookup of https://www.niehs.nih.gov/health/topics/agents/index.cfm. URI lookup does not allow the user to search by page text. }
         \Description{URI lookup in web archives allows user to see every captured version of that webpage in the web archive. One common view is the coffee-stain calendar version used in the web archives like the Wayback Machine, Archive-It, and the Library of Congress. Another version, used in Arquivo.pt, organizes the captures by month in a list rather than a calendar view. Neither interface shows any information about page text, or change of text over time. }
        \label{fig:nihmap}
\end{figure}

Journalists frequently use web archives as evidence.\footnote{\href{https://archive.org/about/news-stories/search?mentions-search=Wayback+Machine}{https://archive.org/about/news-stories/search?mentions-search=Wayback+Machine}} These journalists reference archived copies of webpages not only when the pages are unavailable, but also when the pages have changed over time \cite{frew2022}. As pointed out by Teevan \cite{teevan2007d}, simply saving past versions of webpages without providing proper means for discovery is insufficient to help people find the historical content changes that they are seeking.

\begin{figure}
     \centering
          \begin{subfigure}[b]{0.65\textwidth}
         \centering
         \includegraphics[width=\textwidth]{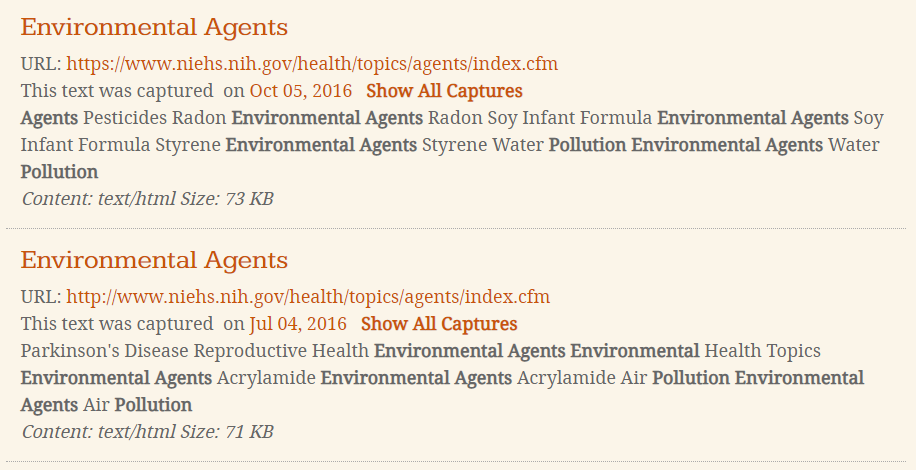}
         \caption{Archive-It collection SERP matching the query in Figure \ref{fig:nih7b} showing title, URI, date, text snippet, metadata, replay link, and link to additional captures. Two captures matching the query are shown individually.}
         \label{fig:nih11b}
     \end{subfigure}
     
     \begin{subfigure}[b]{0.65\textwidth}
         \centering
         \includegraphics[width=\textwidth]{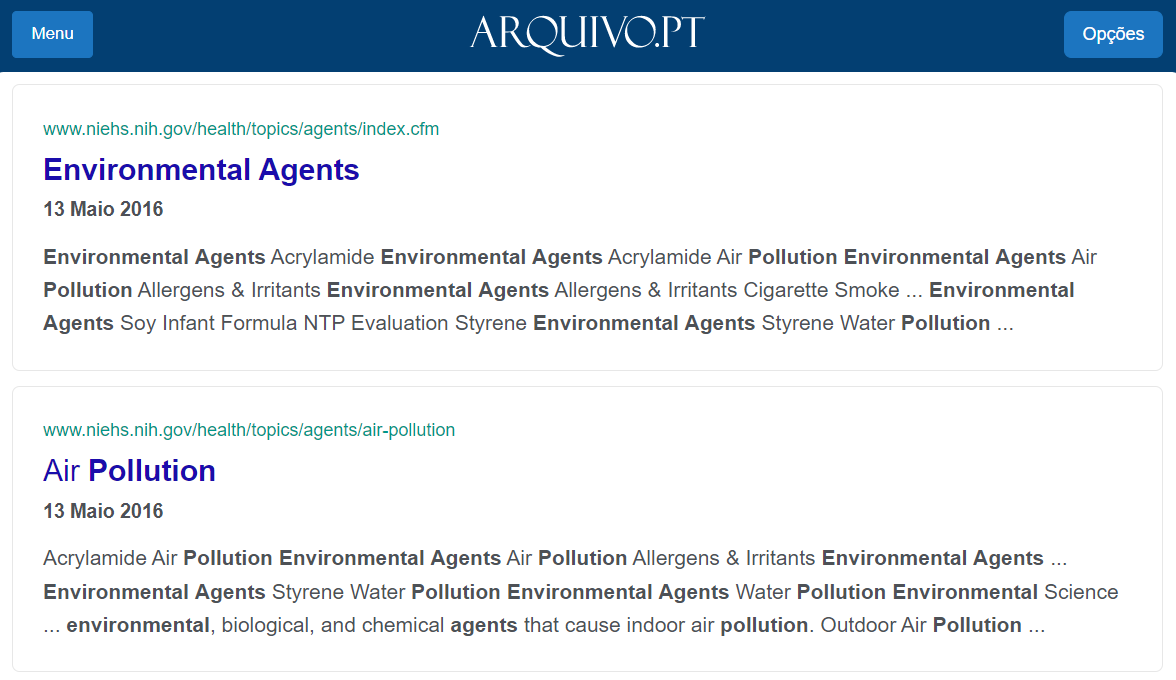}
         \caption{Arquivo.pt SERP matching the query in Figure \ref{fig:nih6b} showing URI, title, date, text snippet, and replay link. }
         \label{fig:nih2b}
     \end{subfigure}
     
     \begin{subfigure}[b]{0.65\textwidth}
         \centering
         \includegraphics[width=\textwidth]{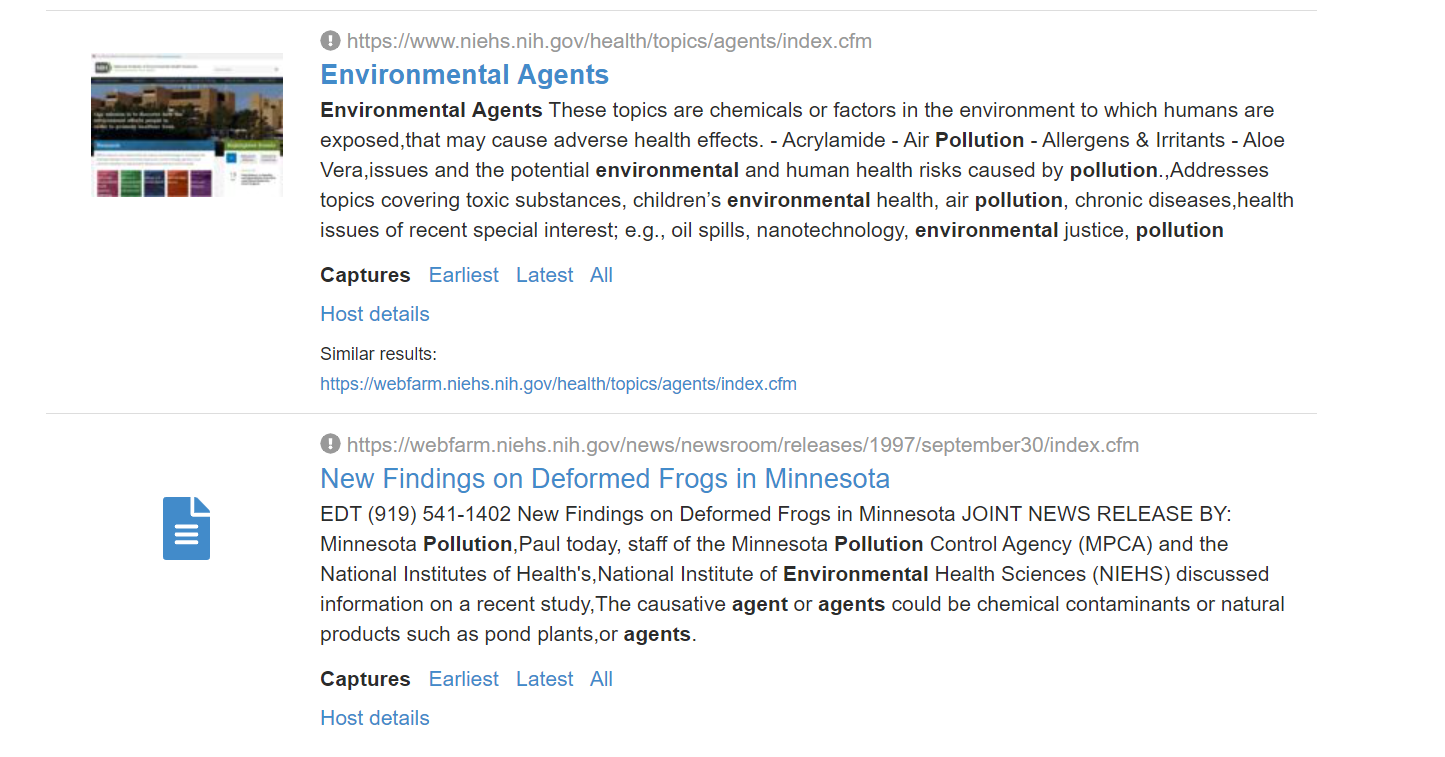}
         \caption{Wayback Machine collection SERP matching the query in Figure \ref{fig:nih8b} showing URI, title, text snippet, screenshot, replay link, and link to additional captures.}
         \label{fig:nih4}
     \end{subfigure}
     
        \caption{None of these three web archive search interfaces group the versions to show the change in the page over time.
        }
         \Description{The Wayback Machine shows that it contains a page matching the search query, but no other temporal information other than the datetime in the replay link. Arquivo.pt shows a page from 2019 matching the query, but does not include captures of that same page from 2016 that should also match the query. Archive-It shows a page matching the search query with temporal information, and that there are additional captures of that page, as well as including each matching capture as its own result.}
        \label{fig:nihserp}
\end{figure} 

\begin{figure*}
     \centering
     \begin{subfigure}[b]{0.3\textwidth}
         \centering
         \includegraphics[height=6cm]{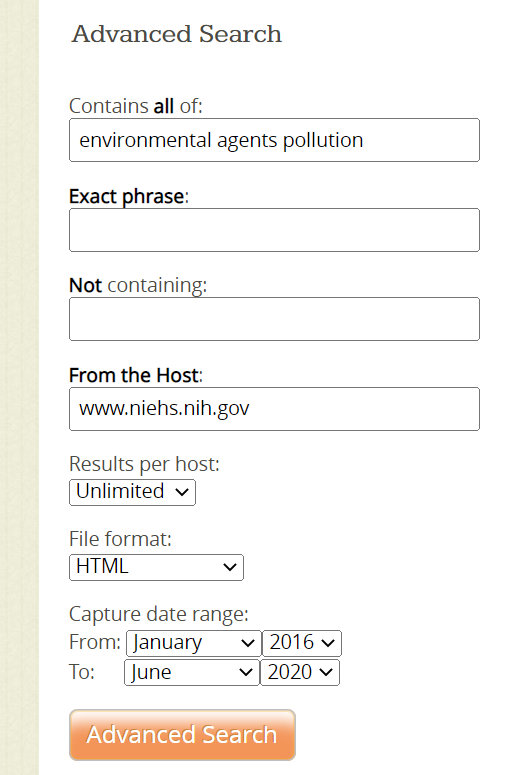}
         \caption{The advanced search box at Archive-It for the National Institutes of Health Web Archive collection 1170 allows for full-text search and filtering by date and domain.}
         \label{fig:nih7b}
     \end{subfigure}
     \hfill
     \begin{subfigure}[b]{0.65\textwidth}
         \centering
         \includegraphics[width=\textwidth]{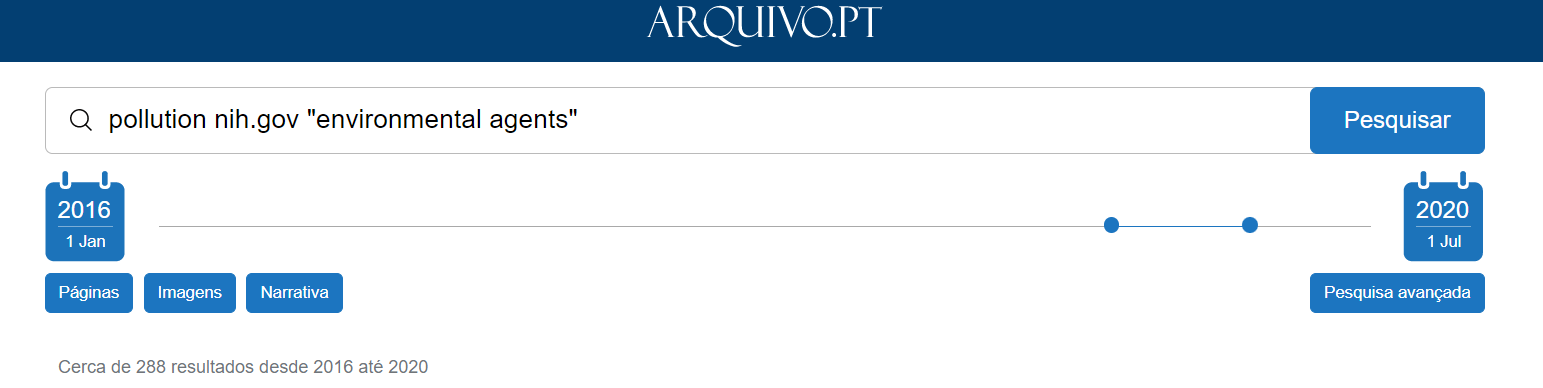}
         \caption{The search box over the entire Portuguese Web archive allows for full-text search and filtering by date.}
         \label{fig:nih6b}
     
         \centering
         \includegraphics[width=\textwidth]{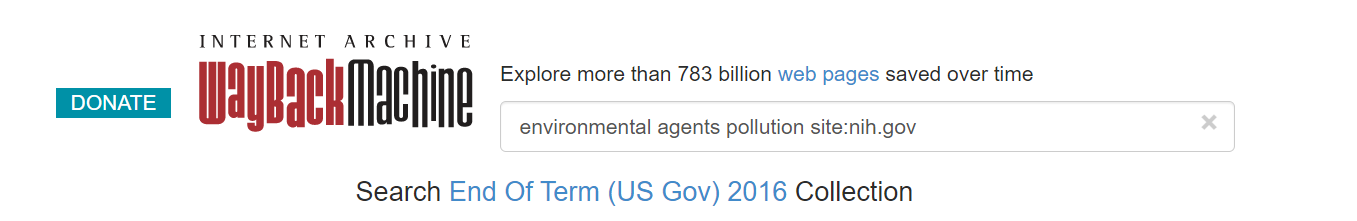}
         \caption{The search box at the Internet Archive's Wayback Machine for the 2016 End of Term Archive collection allows for full-text search and filtering by site, but only over one end of term collection at a time. }
         \label{fig:nih8b}
     \end{subfigure}
        \caption{Some web archives have full-text search with various filtering features over small collections or over the entire web archive. None of the search interfaces allow for users to search for terms that have been removed from webpages.}
         \Description{Three examples of web archive search boxes are shown. While date filtering and domain filtering are helpful, they do not address how to display multiple versions of the same web page that match the search query. These pages may or may not have differences in content, and these interfaces fail to inform the user how the content in the matching versions has changed over time.}
        \label{fig:nihsearchbox}
\end{figure*}

Full-text search has been identified \cite{ras2007web} as a highly requested feature for web archives. However, many web archives only include URI lookup of holdings rather than full-text search, as shown in Figure \ref{fig:nihmap}. 
The Internet Archive's Wayback Machine does provide a search engine, but it is based on metadata, such as page title and domain, rather than full-text indexing. 
Some web archives and individual web archive collections do have full-text search engines. Figure \ref{fig:nihserp} shows the search engine results pages for some of these search engines, and Figure \ref{fig:nihsearchbox} shows the query fields currently available to users in these search engines. None of the current interfaces allow the user to query for webpage changes.

For example, many instances of the term ``pollution" were removed from the \href{https://www.niehs.nih.gov/health/topics/agents/index.cfm}{``Environmental Agents" webpage on nih.gov} between 2016 and 2020, but none of the current web archive search interfaces can help users readily find the exact date and time of the specific change. Figure \ref{fig:nih11b} shows multiple versions of the same webpage as individual search results at Archive-It's National Institutes of Health Web Archive collection.\footnote{\href{https://archive-it.org/collections/1170}{https://archive-it.org/collections/1170}} It is unclear why one result is ranked higher than another, and the current interface does not identify how the versions are different. Figure \ref{fig:nih2b} shows one matching page version at the Portuguese Web Archive,\footnote{\href{https://arquivo.pt/}{https://arquivo.pt/}} but this web archive actually holds additional matching versions of this webpage that are not included in the search results. Figure \ref{fig:nih4} shows a search of the 2016 End of Term archive on the Internet Archive's Wayback Machine,\footnote{\href{https://web.archive.org/EndOfTerm2016WebCrawls/search/}{https://web.archive.org/EndOfTerm2016WebCrawls/search/}} yet the version linked in the search result is from January 24, 2017, after inauguration day. Later, we will present a more effective search results page for changes in webpages, showing that the deletion occurred between February and March 2017. 

In this work, we contribute a temporal search engine architecture and interface for web archive collections that allows users to search for changes on webpages and view those changes in context. The main contribution of the search engine architecture is the calculation of changed terms between mementos. The main contributions of the search interface are a search engine that can query for changes, a search engine results page that shows multiple versions of a page as one result in a meaningful way, a sliding difference viewer that allows users to examine a page's text change over multiple time periods, and a difference animation that allows users to see changes in context.

The change text search engine backend was evaluated with the Environmental Data and Governance Initiative's (EDGI) federal environmental webpages data set \cite{nost2021}. Users can view the webpage changes on a search engine results page snippet. The indexing process can identify additional terms beyond the original terms identified in the original EDGI study, and the temporal granularity of the deletions can be increased compared to the original four-year window from the original study.

\section{Background and Related Work} 

The Memento Protocol \cite{van2013http} is the standard HTTP content negotiation protocol for web archives. Users request a specific webpage according to its URI on the live web, also known as its URI-R, along with a datetime. Memento-compliant web archive servers will query for and return the archived version of the webpage closest to the requested datetime. Each archived version of a webpage can be accessed directly with its own URI, also known as a URI-M. The archived versions of webpages are also known as \emph{mementos}. Finally, a listing of all archived versions of a webpage on a specific server can be queried. This listing is called a \emph{TimeMap}, or URI-T. Mementos are commonly referred to as being a part of the past web, while a still-working URI-R would belong to the live web. There are multiple web archives, including national web archives like the Portuguese Web Archive, subscription-based web archives for organizations like Archive-It, and more comprehensive web archives like the Internet Archive's Wayback Machine. While the Memento Protocol is standard in web archives, there is no standard level of support for full-text search. Web archives contain multiple versions of pages over time, yet existing full-text temporal search engines for web archives do not successfully address how to present multiple versions of the same page.

Web archives contain a vast amount of untapped potential for analyzing change over time. Webpages change at different rates according to features such as their domain and the depth of the URL \cite{adar2009web}. Existing data sets of webpage crawls commonly used for academic purposes, such as ClueWeb \cite{overwijk2022clueweb22} and Common Crawl,\footnote{\href{https://commoncrawl.org/}{https://commoncrawl.org/}} aim to collect snapshots of a large amount of unique URLs. Since these data sets did not aim to collect multiple versions of the same page, there is no guarantee of any particular page being crawled regularly, which would hinder the ability to analyze change over time. Additionally, in order for an analysis of webpage change to be of consequence, the changes on those webpages need to be meaningful. Clearly, examining topic-based change over time using web archives is limited because searching for pages with specific changes is not currently possible. 

In order to capitalize on using web archives to examine change over time, temporal search interfaces must be extended to support searching for term and phrase changes. These search interfaces will also need to effectively show the changes to the users. Finally, the search tool will need to be able to find meaningful changes in a data set of webpages relevant to existing digital humanities research. 



\subsection{Temporal Search Engines for Web Archives}

Full-text search is not a feature available in every web archive.
Some web archives, such as the Portuguese Web Archive\cite{melo2016architecture}, the UK Web Archive,\footnote{\href{https://www.webarchive.org.uk/ukwa/}{https://www.webarchive.org.uk/ukwa/}} and collections in Archive-It,\footnote{\href{https://archive-it.org/}{https://archive-it.org/}} do support full-text search. Each of these systems is independent; there is no full-text search engine that can search all of the versions of a webpage across multiple web archives. One of the major challenges in creating a temporal search engine for web archives is that each archived web page may have multiple versions. Ready-made information retrieval systems for versioned document collections do not exist. Indexing, presenting search results, and displaying changes in context all need to be addressed when creating such a system. 

How versioned document collections are indexed is vital to discovery of changes between versions. Berberich et al. \cite{Berberich2007} developed a temporal coalescing framework that orders all versions of a document over time and then assigns each document a validity range rather than a single datetime. This allowed for the amount of document change to be quantified between versions. While documents with a certain similarity threshold were combined in the index to save space, the separate document versions were not combined in the search engine results page, and searching by change was not possible. While temporal date ranges were not implemented when Berberich developed the validity range framework, date range fields are now a standard part of Apache Lucene \cite{smiley2015apache}, the premiere open-source search backend.\footnote{\href{https://lucene.apache.org/}{https://lucene.apache.org/}} 

A few temporal search engines have investigated how to present multiple versions of a page in the search results. Melo et al. \cite{melo2016architecture} added a backend parameter to the Portuguese Web Archive search system to limit the number of versions displayed on a search results page. Kiesel et al. \cite{Kiesel2018} performed a qualitative evaluation on their personal web archiving localized search system, and found that including every version of a web page introduced clutter into the search results pages that affected usability. Major \cite{major2022} also identified repeated captures of the same URI in search results as problematic. On the other hand, Jackson et al. \cite{Jackson2016} chose to include every relevant version of a web page in their search results page ordered by time. This was because grouping web page versions would hide change over time. In all of these systems, searching by change is not possible, and the only way to view the changes between versions is by manual inspection.

\subsection{Showing Change Between Document Versions}

Temporal search engines for web archives do not incorporate the changes between versions of webpages into their results, but some other tools do exist to help users find and view changes on known pages. Sherratt et al. \cite{Sherratt2022} created a tool to help users find changes in a specific web page's version history as a part of the GLAM Workbench web archives Jupyter notebook collection. Because this tool does not index any web page content, each URL must be searched individually. The tool also only includes a linear search option, which hinders its speed. The results highlight the query term in context, but not in context of the previous versions of the page. Another tool, WikiBlame,\footnote{\href{http://wikipedia.ramselehof.de/wikiblame.php}{http://wikipedia.ramselehof.de/wikiblame.php}} allows for users to search for changes in a specific Wikipedia page. While this tool does allow for binary search, it does not index any page content, so only one page can be searched at a time rather than an entire group of related pages. Wikipedia includes a differences tool as part of its version history viewer,\footnote{\href{https://en.wikipedia.org/wiki/Help:Diff}{https://en.wikipedia.org/wiki/Help:Diff}} allowing users to view changes in a static context.

\begin{figure*}
     \centering
     \begin{subfigure}[b]{\textwidth}
         \centering
         \includegraphics[width=\textwidth]{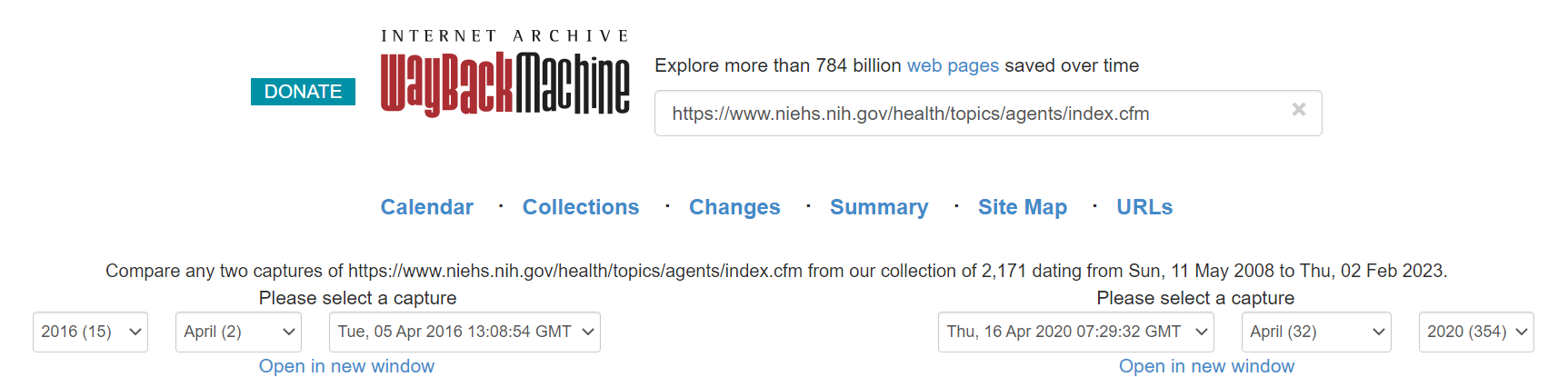}
         \caption{The Wayback Machine Changes tool requires the user to know the date and time of both versions of the page in order to create a comparison. }
         \label{fig:nih9}
     \end{subfigure}
     \hfill
     \begin{subfigure}[b]{0.7\textwidth}
         \centering
         \includegraphics[width=\textwidth]{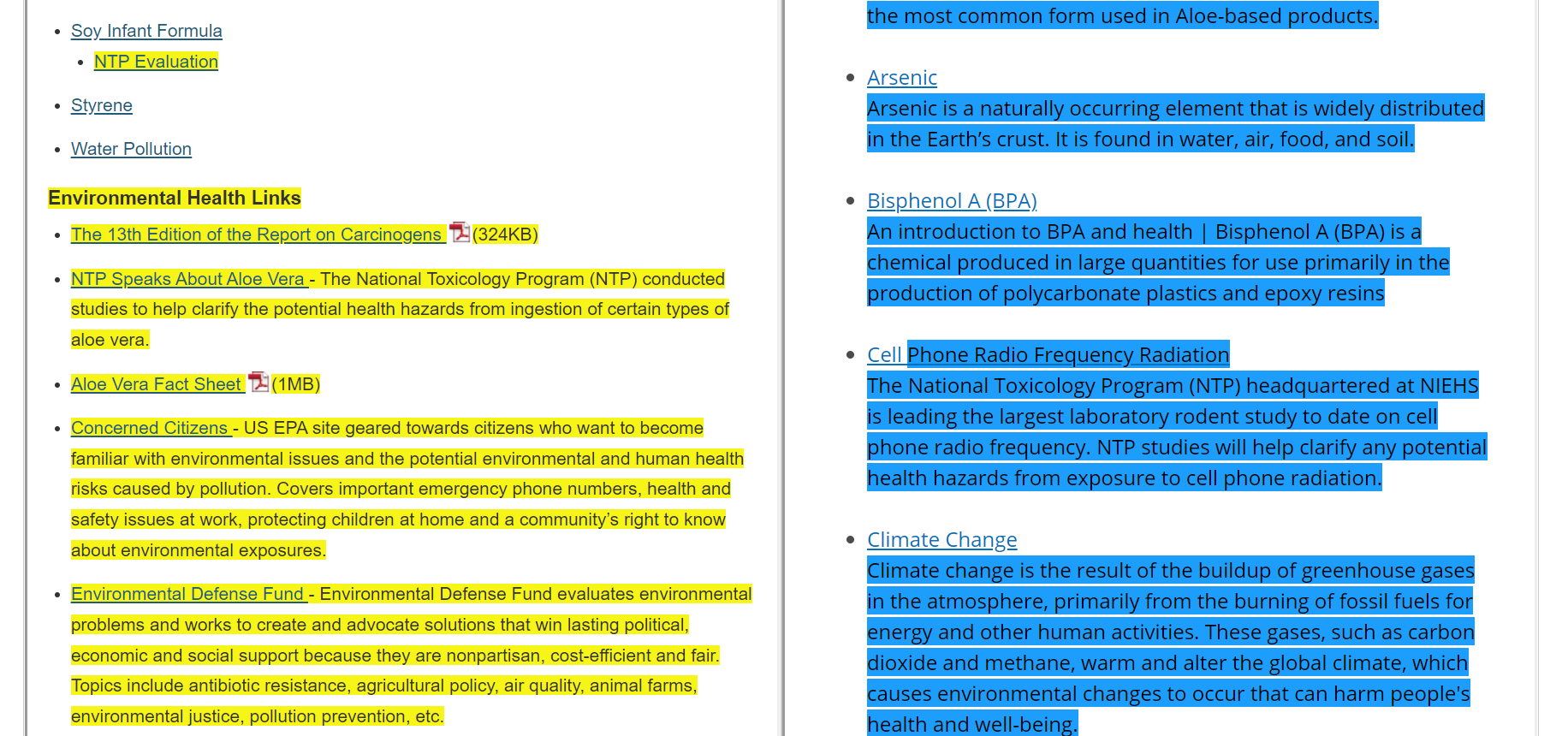}
         \caption{The Wayback Machine Changes tool helps users examine additions (blue) and deletions (yellow). The term pollution, on the left in yellow, was removed.}
         \label{fig:nih10}
     \end{subfigure}
        \caption{The Internet Archive's Wayback Machine Changes tool shows users the differences between two captures. It only works for captures at this one web archive, and there is no way to search the changes to find the dates and times. }
         \Description{On the left, the version of the Environmental Agents nih.gov page from 2016 shows a content section about pollution. The section is highlighted in yellow, indicating it was deleted. The right panel shows the page in 2020, without the content about pollution. The exact time and date of the content removal between 2016 and 2020 is not available.}
        \label{fig:nihdiffwb}
\end{figure*}

The Internet Archive Wayback Machine allows users to compare two different versions of a webpage using its Changes tool.\footnote{\href{https://web.archive.org/web/changes/}{https://web.archive.org/web/changes/}} This tool, shown in Figure \ref{fig:nihdiffwb}, was built upon the Environmental Data and Governance Initiative's Web-Monitoring-Diff suite.\footnote{\href{https://github.com/edgi-govdata-archiving/web-monitoring-diff}{https://github.com/edgi-govdata-archiving/web-monitoring-diff}} This service is not currently connected with any tool that allows users to search for changes across versions, and only lets users compare versions from the Internet Archive. There is currently no version of the Changes tool that allows users to compare versions of webpages that exist across multiple archives. The Changes tool is only able to compare two versions of a webpage at one time, and the comparison is currently presented in a side by side static context.

Other types of versioned document collections besides web archives have led to tools that allow for comparison for multiple document versions. For example, Henley et al. \cite{Henley2016} developed a tool called Yestercode that allows programmers to use a slider to navigate between different versions of their code and display the differences between consecutive versions. The collaborative document writing tool DocuViz \cite{wang2015} aims to help users visualize how documents evolved, and Perez-Messina et al. \cite{perez2018} developed a tool to visualize the origin of text segments in collaborative documents. 

Careful design choices are necessary to help users spot changes. Preattentive processing is a core technique employed during design to aid in rapid detection of changes, such as in estimation situations \cite{healey1996}.
One system that animates changes in text as a complement to displaying the changes side-by-side is Diffamation \cite{chevalier2010}. This system shows all text changes between two versions of a document in parallel animation with navigation in order to help the user understand where, what, and how much text has changed. Another system that uses text animation to show differences between document versions is SlideDiff \cite{denoue18}. 
SlideDiff shows changes to text and media in slide presentation versions, even showing a simulated mouse cursor in order to draw the user's attention to the position of each of the changes and help the user infer the intent of the editor due to the animation appearing more life-like. These goals are also applicable to viewing and understanding edits on webpages.
Jatowt et al. \cite{Jatowt2006} created the Journey to the Past framework that included a browser that allowed users to search for changes in a web page across web archives and animate the differences that matched the query terms. The interface for this browser is similar to video replay, with buttons to help the user navigate between versions and control the animation replay. Because this framework does not include indexing, searching for a changed term or phrase must be done one page at a time, rather than at the collection level. Adar et al. \cite{Adar2008} developed the Zoetrope system to allow users to search individual web page versions for terms and show the differences over time in a stop-motion style of animation. This system relied on local crawling, and the queries relied on closest matches in the DOM rather than indexing; integrating full-text search with Zoetrope's visualization capabilities was identified as future work.

\subsection{Changes on Federal Websites}

The archival of federal websites, especially at the end of a president's term, is an important task undertaken by multiple organizations. The End of Term Web Archive is created through a partnership between five organizations, including the Internet Archive and the Library of Congress \cite{seneca2012takes, phillips2017end}. This web archive includes a full-text search feature,\footnote{\href{http://eot.us.archive.org/search/}{http://eot.us.archive.org/search/}} but each end of term crawl includes only one capture of each web page. Phillips et al. \cite{phillips2016exploratory} compared the 2008 and 2012 end of term collections to identify changes in crawl dates and webpage addresses, but individual terms were not analyzed. Nost et al. \cite{nost2021}, on behalf of the Environmental Data and Governance Initiative (EDGI), compared the change in 56 pre-chosen environmental terms and phrases between 2016 and 2020 using the web archive holdings at the Internet Archive. They identified a list of approximately 10,000 webpages that had versions in both 2016 and 2020. The EDGI analysis focused on highlighting collection-level trends, rather than creating an interface to allow users to navigate to changes in individual pages. Both the 2008/2012 and 2016/2020 analyses relied on changes between versions captured four years apart, but many additional versions of these pages exist across multiple web archives that have not been analyzed for changes. 

\section{Design Motivation}

Developing a mental model of web archives and learning how to navigate the past web are not trivial tasks for many users. As recently as 2019, Abrams et al. \cite{abrams2019sowing} found that users had difficulty distinguishing between whether they were on the live web or past web. They also found that users' lack of understanding of web archives hindered their success as much as an ineffective user interface. Full-text search is therefore an advanced feature that will only benefit users with a strong understanding of the past web.

\begin{figure}[h]
  \centering
  \includegraphics[width=0.8\linewidth, trim= 0 0 0 25, clip]{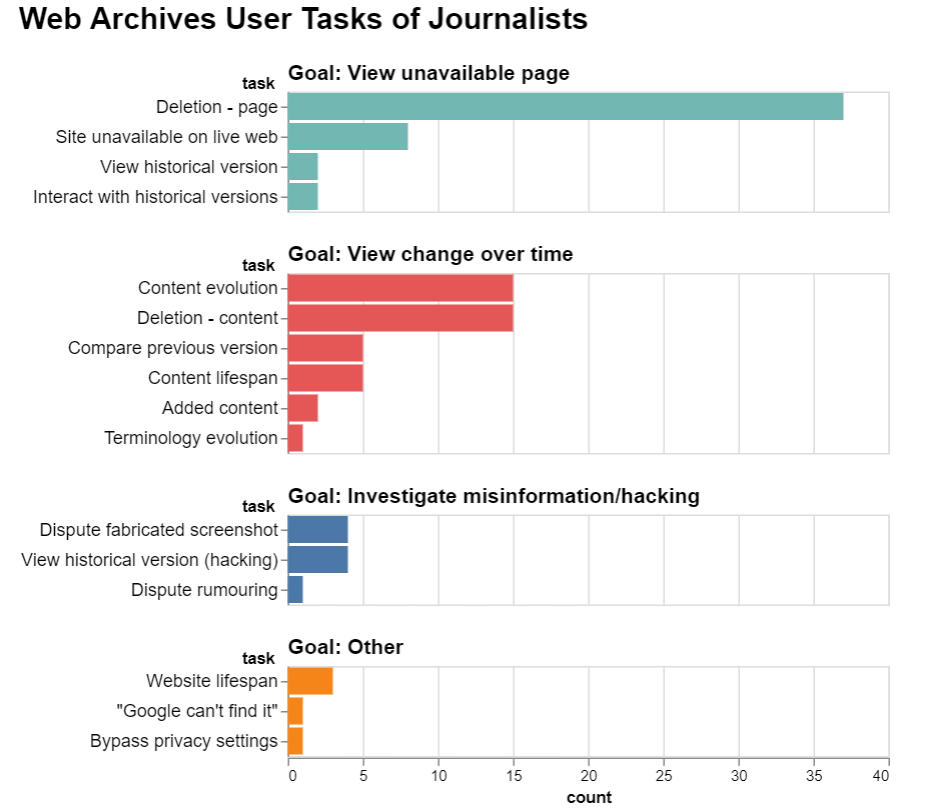}
  \caption{The two most common goals for journalists who use web archives as evidence in their articles is to view unavailable pages and to view page content change over time.}
  \Description{Journalists use web archives to support different goals, such as viewing unavailable pages, viewing change over time, and investigating misinformation. Users who want to view change over time are interested in term additions, term deletions, and content lifespan. }
  \label{fig:journ}
\end{figure}

One group of candidate users for full-text search in web archives are journalists. There has been no prior analysis of journalists' use of web archives \cite{Ogden21}. Teevan conducted a prior study analyzing why general users want to interact with lost webpages \cite{teevan2007d}. We conducted a brief naturalistic study using similar methodology as Teevan in order to categorize the tasks of journalists using web archives, in order to determine if full text search could benefit them.

We searched Google News for the phrase ``Wayback Machine" using the GNews Python API,\footnote{\href{https://news.google.com/}{https://news.google.com/}, \href{https://github.com/ranahaani/GNews/}{https://github.com/ranahaani/GNews/}} and collected 500 articles matching this query on a biweekly basis from May to July, 2022. Out of the 500 collected articles, 106 of them used web archives as evidence. We manually categorized these articles by user task. The journalists stated how they used web archives in their articles, which is how we created the task codes. Similar to Teevan, one author coded the dataset. As shown in Figure \ref{fig:journ}, journalists use web archives to view unavailable pages, but they also frequently use web archives to view change over time. These users wanted to view term and phrase additions, deletions, and the associated content lifespan 
\cite{frew2022}.

Currently, finding additions and deletions of terms and phrases is only possible by manually inspecting multiple versions of a webpage. Since journalists do have a strong mental model of web archives, their tasks would be made easier by a search engine that allowed them to find the changes in content over time. 

While journalists are the only group of users we studied, they are not the only users who would benefit from such a tool. The IIPC Access Working Group \cite{iipc2006} identified additional professional users including researchers studying change over time in computational and humanities contexts, professionals investigating data evolution such as in tourism or real estate, and lawyers in civil trademark cases and patent cases.

Instead of considering how to display multiple versions of webpages in search results as a hindrance, finding and showing the changes between the versions should be one of the main features of a temporal search engine for web archives. 

\section{Architecture}

\begin{figure*}[h]
  \centering
  \includegraphics[width=12cm]{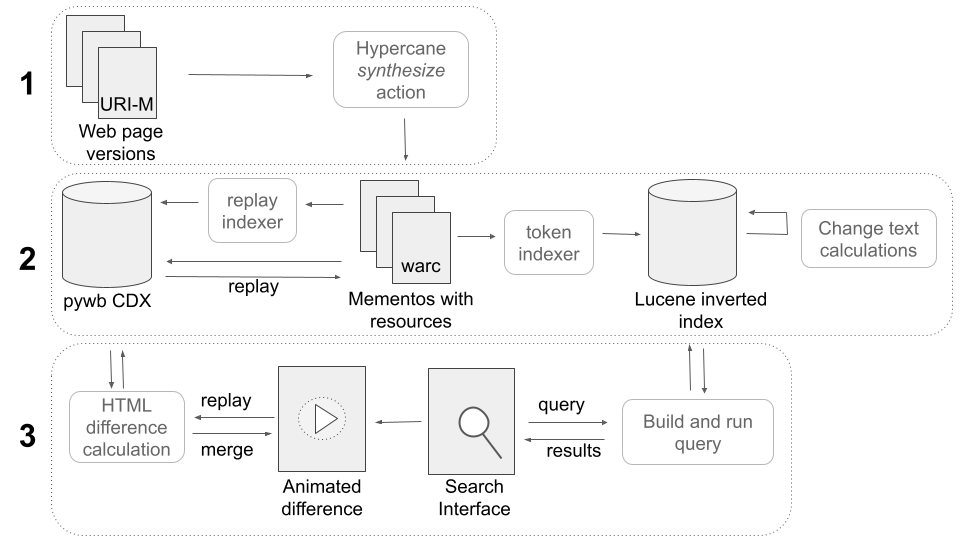}
  \caption{The architecture of the change text search engine. Level 1 consists of document acquisition. Level 2 consists of the documents and indices. Level 3 consists of the user interface.}
  \Description{In Level 1, documents are acquired using Hypercane. In Level 2, WARC documents are indexed for both tokens and replay. Additional processing is done on the Lucene index to calculate change text. In Level 3, the Lucene index is queried for search engine results. The PyWB index is queried for difference animation.}
  \label{fig:arch}
\end{figure*}

The architecture for the change text search engine consists of three levels, as shown in Figure \ref{fig:arch}. First, webpages with multiple versions must be acquired. Next, the webpages are indexed for both text content and replay. The changes in the text content are calculated after the initial indexing. Finally, the user interface for a search engine provides the user with a way to discover the changes in the webpages, and replay those changes in context.

\subsection{Document Acquisition}

The documents for the change text search engine need to be in WARC format. While users can replay public web archives' holdings, they cannot access the original WARC files. Indexing these public holdings into the change text search engine is therefore not possible without a tool that can turn a URI-M into a WARC file. Hypercane \cite{Jones21} is a tool for interacting with web archive collections. One feature of the Hypercane tool is to \emph{synthesize} a WARC from a given URI-M. The WARC file output by Hypercane includes the original HTML of the web page in raw form, along with all available embedded resources. The HTML is necessary for creating an inverted index of the changes in page content, while the embedded resources are necessary for merged replay showing the differences between page versions in context. Users and organizations with existing WARC collections would not need to perform any additional document acquisition. 

\subsection{Indexing}

The document collection must be indexed in order to support querying and replay. The first index generated is the Apache Lucene index that contains the tokenized content of the pages, along with additional page metadata. The UK Web Archive WARC indexer \cite{ukwai} is an existing tool that can properly index WARCs for Lucene. Apache Solr provides an interface for working with Lucene.\footnote{\href{https://solr.apache.org/}{https://solr.apache.org/}} We extended the Solr schema originating from SolrWayback \cite{solrwb} to support new change text fields (for sets of added and deleted terms) prior to indexing. We wrote a change text calculation script to populate these new fields. This Java code interfaces directly with the Lucene index to calculate the sets of added and deleted terms as well as the document temporal validity ranges as defined by Berberich et al. \cite{Berberich2007}. Documents are grouped according to their canonicalized URL, which is an existing field populated during the initial indexing.

The second index generated is the PyWB \cite{pywb} CDX that contains the information needed to replay the pages. Since PyWB's default behavior is to make a copy of all indexed WARCs in a single folder, this behavior was turned off in order to reduce disk space usage. The folders containing the WARCs can simply be added into the PyWB configuration file to enable successful replay without duplication.

\subsection{User Interface}

\begin{figure}[h]
  \centering
  \includegraphics[width=\linewidth]{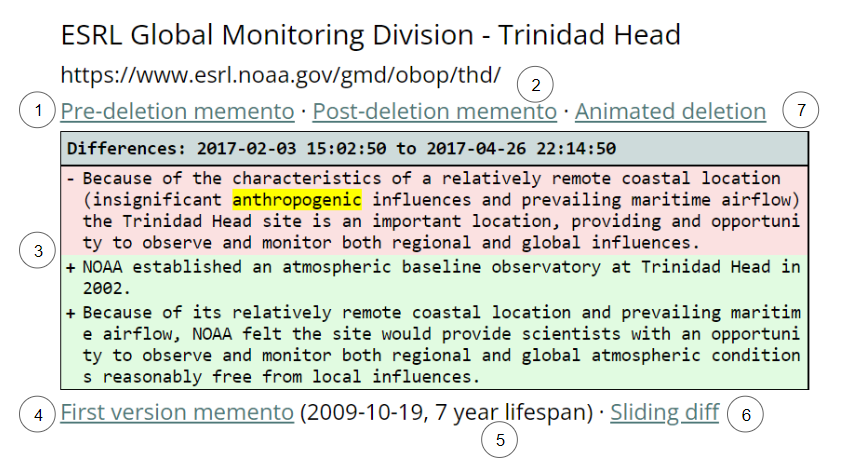}
  \caption{Change text search interface. 1, 2, and 4: individual replay links to page mementos; 3: the diff between the pre and post deletion versions; 5: content lifespan calculation; 6: the link to the sliding diff viewer across all indexed versions of the page; 7: the link to the deletion animation}
  \Description{The search engine results page includes 7 main parts. Three versions, the pre-deletion, post-deletion, and addition drive the calculations. The difference between the pre and post deletion is shown, with the search term highlighted.}
  \label{fig:ui}
\end{figure}

A user can interact with the change text term index through a search interface,
which we built 
using Solarium,\footnote{\href{https://github.com/solariumphp/solarium}{https://github.com/solariumphp/solarium}} a PHP Solr interface. A user may query for a deleted term, a deleted phrase, an added term, or an added phrase. Deleted terms and phrases can be detected whether the term is fully, or only partially, removed from the webpage. The type of change, along with the term or phrase, is transformed into a valid Lucene query over the appropriate change text fields, and then the search results are returned and displayed in a search engine results page, shown in Figure \ref{fig:ui}.

\begin{figure*}[h]
  \centering
  \includegraphics[width=\linewidth]{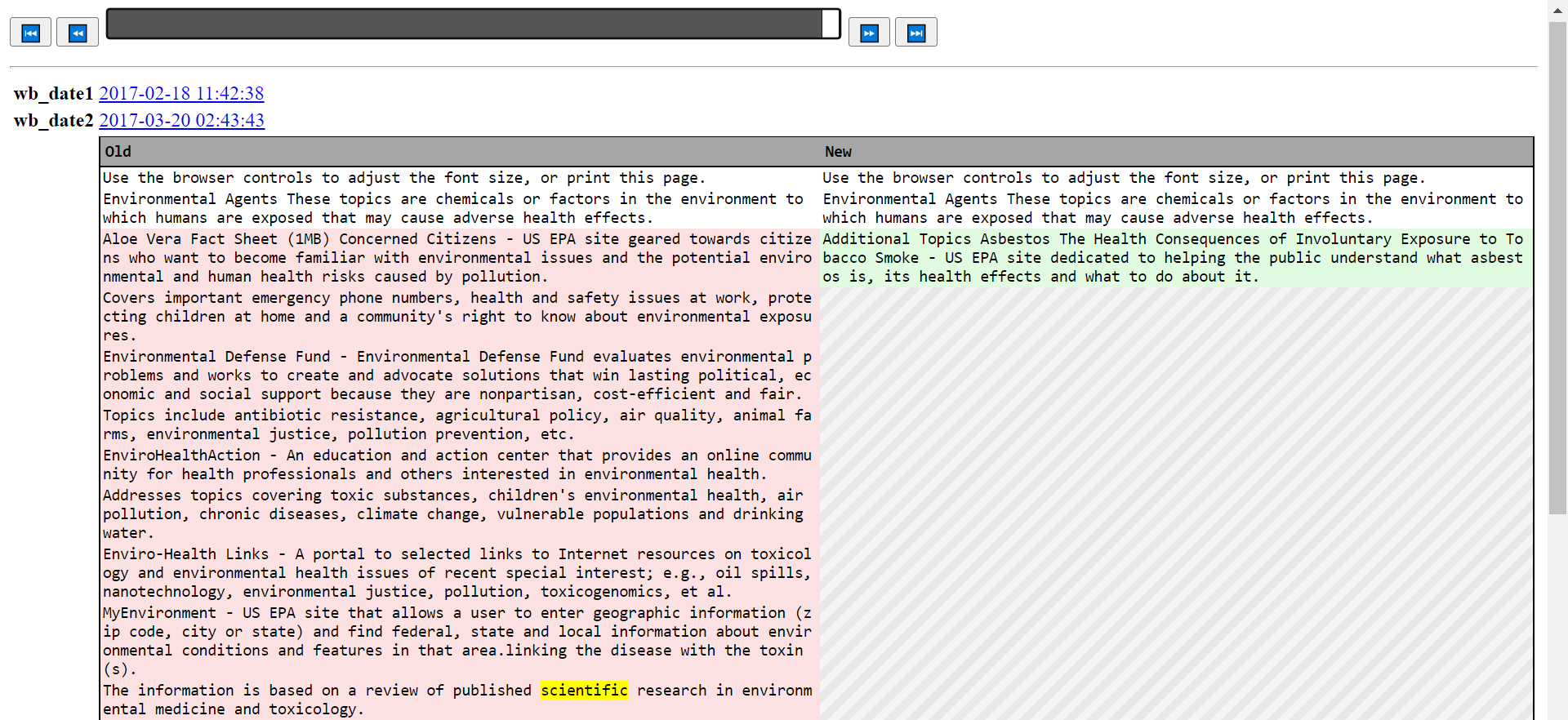}
  \caption{The sliding difference viewer shows the term `scientific' was deleted from the page https://www.niehs.nih.gov/health/topics/agents/index.cfm in 2017, along with the context of the deletion.}
  \Description{The sliding difference viewer has navigation buttons that allow the user to skip identical versions, indicated by the fast forward and rewind buttons. There are also buttons to navigate to the first and last versions, which indicate the addition and the deletion of the term, indicated by the skip to end and skip to beginning buttons. Finally, there is a sliding bar that the user can drag to navigate between consecutive versions.}
  \label{fig:slidediff}
\end{figure*}

\clearpage

\begin{figure}[hb]
  \centering
  \includegraphics[width=0.6\linewidth]{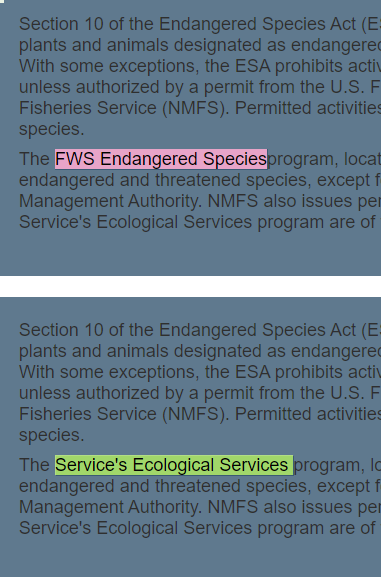}
  \caption{The animation shows the deletion of the phrase ``endangered species" on the page http://www.fws.gov/ENDANGERED/permits/index.html.}
  \Description{Before the animation begins, the deletion is highlighted in red. Each letter is deleted one by one, and if there are enough words in the deleted content, whole words will be deleted starting with the fourth word to speed up the animation. Next, the addition is shown highlighted in green. Each letter is animated in the addition one by one. Multiple deletions are shown with a jump to each block of changing page content.}
  \label{fig:anim}
\end{figure}

The change-text search interface allows the user to understand the differences between multiple versions of the same page, unlike the current web archive interfaces shown in Figure \ref{fig:nihserp}. Each search engine result shows information about when the term was added and/or removed, along with a snippet using hues showing the difference between two consecutive versions most relevant to the query. The search engine result includes links to replay these versions individually via standard PyWB replay, a link to compare all page versions between the addition and deletion as shown in Figure \ref{fig:slidediff}, and a link to replay the pre- and post-deletion versions simultaneously as shown in Figure \ref{fig:anim}. In the first tool, the sliding difference tool, the user can skip past identical mementos with the fast forward and rewind buttons. We created this sliding difference viewer by using the plaintext indexed content from Lucene. The viewer functions using a Solarium query, the PHP difference library \cite{cherng2023},
and some additional JavaScript. The second tool, the dual replay tool, shows an animation of the difference in context. This tool is shown in action at \href{https://youtu.be/qHSVvcubuYo}{https://youtu.be/qHSVvcubuYo}. The differences in the animation use hues. Both the animation itself along with the highlighted text colors lend themselves to pre-attentive processing of the changes. The animation jumps to each change in turn, in contrast to the static Changes Tool on the Wayback Machine. The animation is also meant to give the illusion of the changes happening in real time. In order to create this animation, we used EDGI's Python HTML difference library to calculate and combine the differences in a static context, and then extended the code to generate the HTML and JavaScript to animate the merged pages. We support successful difference calculation between page versions that originated from different web archives by using bannerless replay with PyWB. To draw the user's attention to each change, the page jumps to each change one-by-one, animates one change at a time, and pauses before jumping to the next change.

\begin{figure*}
     \centering
     \begin{subfigure}[b]{\textwidth}
         \centering
         \includegraphics[width=0.8\textwidth]{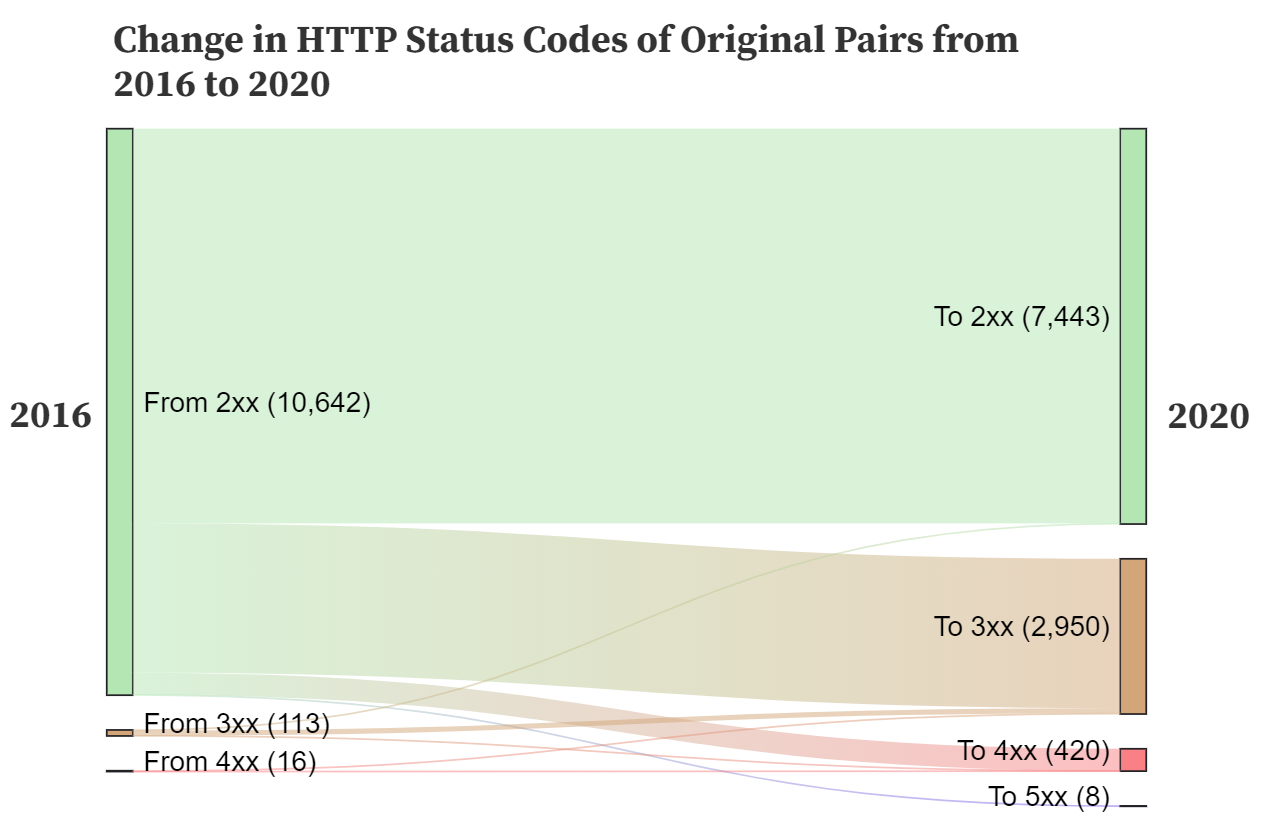}
         \caption{About 75\% of the paired mementos identified by EDGI have successful (200) HTTP status codes in both 2016 and 2020.  }
         \label{fig:codeorig}
     \end{subfigure}
     
     \begin{subfigure}[b]{\textwidth}
         \centering
         \includegraphics[width=0.8\textwidth]{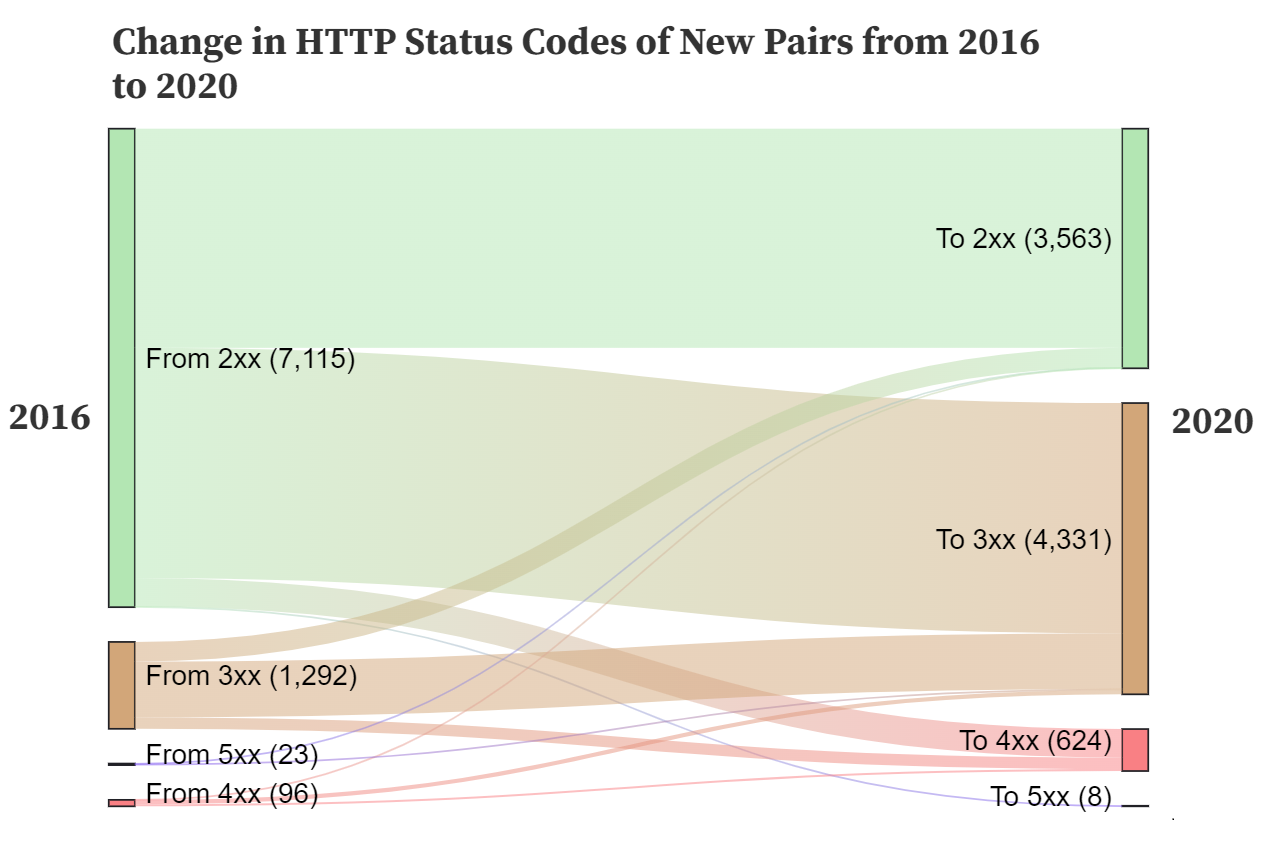}
         \caption{In the new pairing set, an additional 3,500 paired mementos have successful (200) HTTP status codes in both 2016 and 2020.}
         \label{fig:codenew}
     \end{subfigure}
        \caption{Out of about 40,000 seed URI-Rs in the EDGI data set, approximately 11,000 have successful (200) HTTP status codes in both 2016 and 2020. We found an additional 3,500 mementos from multiple web archives with successful status codes. }
         \Description{The original EDGI set contained only mementos from the Internet Archive, while the new set contains mementos from multiple web archives. Most of the original paired mementos have a 2016 status code of 200. Of the pairs that do not have a 200 status code in 2020, many have a 300 status code or 400 status code, and just a few have a 500 status code. There are more non-successful HTTP status codes in the new pairing set than in the original set calculated by EDGI. There are more mementos from 2020 that result in redirect status codes than successful status codes.}
        \label{fig:codecomb}
\end{figure*}

\section{Evaluation}

\subsection{Finding changes on federal webpages}

Federal webpages with changes between 2016 and 2020, as calculated by EDGI, form an excellent data set for evaluation of a change text search engine. The EDGI data set consists of about 40,000 web page addresses. Approximately 10,000 of these web pages have versions in 2016 and 2020 at the Internet Archive. First, we expanded the data set to include more of the original pages by considering additional web archives. Next, we filtered the results so that only pages with successful HTTP status codes (HTTP 200 OK) are kept for indexing. Finally, we prioritized pages with known term changes, and we identified additional versions of these pages inside of the four-year window with salient changes for indexing.


EDGI examined each monitored page to determine if a capture from both the first half of 2016 and the first half of 2020 existed, which they defined as a paired-page sample. Since only about 10,000 pages of the 40,000 monitored pages had paired mementos at the Internet Archive, there are about 30,000 pages to examine for paired mementos using other web archives. We generated a TimeMap for each of these 30,000 web page addresses using MemGator \cite{alam2016}, a memento aggregation service. Examining the TimeMaps, about 8,500 of the 30,000 pages do have newly found paired mementos. Interestingly, about half of these new pairs exist at the Internet Archive. It is possible that these mementos were added after some kind of delay or embargo period. The other pairs directly rely on web archives besides the Internet Archive for at least one of the mementos in the 2016/2020 pair.

In the original set of approximately 10,000 paired mementos, about 75\% of the pairs have successful HTTP status codes for both 2016 and 2020, as shown in Figure \ref{fig:codeorig}. In the additional set of approximately 8,500 paired mementos, about 38\% of the pairs have successful HTTP status codes for both 2016 and 2020, as shown in Figure \ref{fig:codenew}. In all, there are about 11,000 pairs of 2016/2020 mementos that are candidates for indexing. EDGI collected their data in a way that minimized non-successful HTTP status codes by using the Internet Archive's CDX API.\footnote{\href{https://github.com/internetarchive/wayback/tree/master/wayback-cdx-server}{https://github.com/internetarchive/wayback/tree/master/wayback-cdx-server}} For the new pair set, other web archives do not have public CDX APIs and TimeMaps do not have any status code information included, which is what led to more pages in the new pair set having non-successful HTTP status codes.

\begin{table}[h!]
\centering
\begin{tabular}{ |c|c|c|  }
 \hline
 Web Archive & \% (2016) & \% (2020) \\
 \hline
 webarchive.loc.gov & 59.81 & 60.93 \\
web.archive.org & 34.41 & 71.15 \\
wayback.archive-it.org & 12.41 & 10.36 \\
arquivo.pt & 5.95 & 3.59 \\
perma.cc & 0.39 & 0.39 \\
web.archive.org.au & 0.11 & 0.08\\
swap.stanford.edu & 0.11 & 0.06 \\
wayback.vefsafn.is & 0.11 & 0.20 \\
waext.banq.qc.ca & 0.08 & 0.08 \\
www.webarchive.org.uk & 0.03 & 0.08 \\
archive.md & 0.03 & 0.06 \\
 \hline
 \end{tabular}
\caption{Web archives with mementos in the first half of 2016 and 2020 for the 3,563 pages in the new pairing set with 200-to-200 status codes, as shown in the top right of Figure \ref{fig:codenew}. }
\label{table:wa2016}
\end{table}

\begin{figure*}
     \centering
     \begin{subfigure}[b]{0.8\textwidth}
         \centering
         \includegraphics[width=\textwidth]{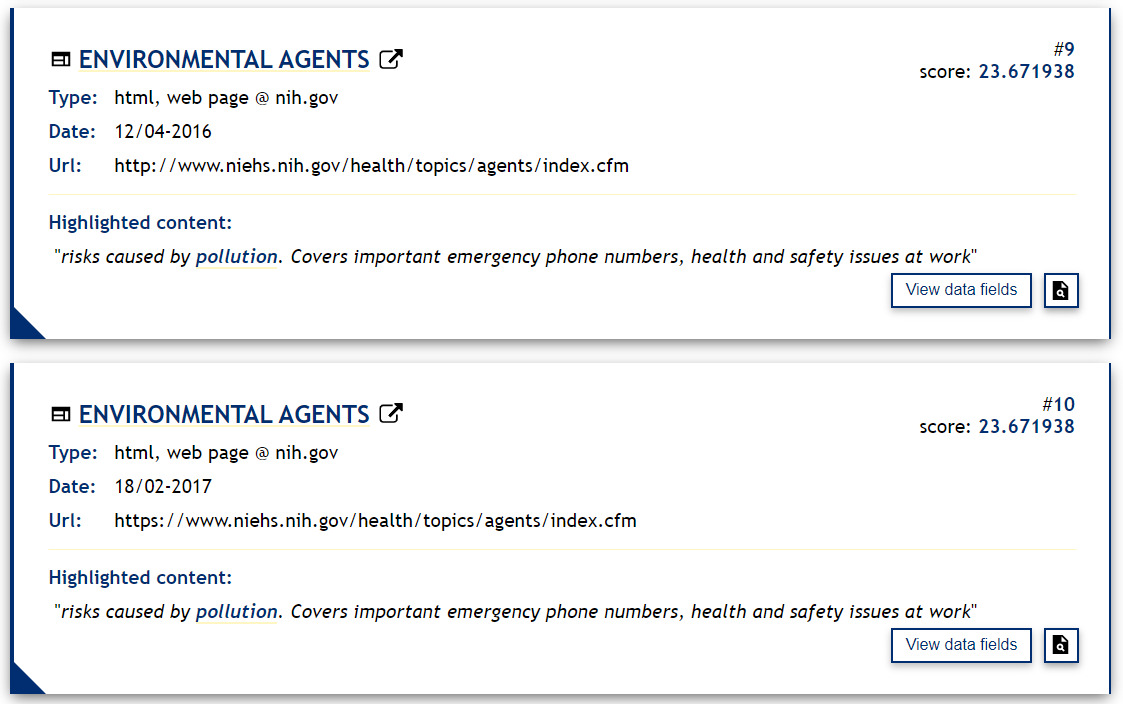}
         \caption{SolrWayback SERP}
         \label{fig:vssolrold}
     \end{subfigure}
     
     \begin{subfigure}[b]{0.8\textwidth}
         \centering
         \includegraphics[width=\textwidth]{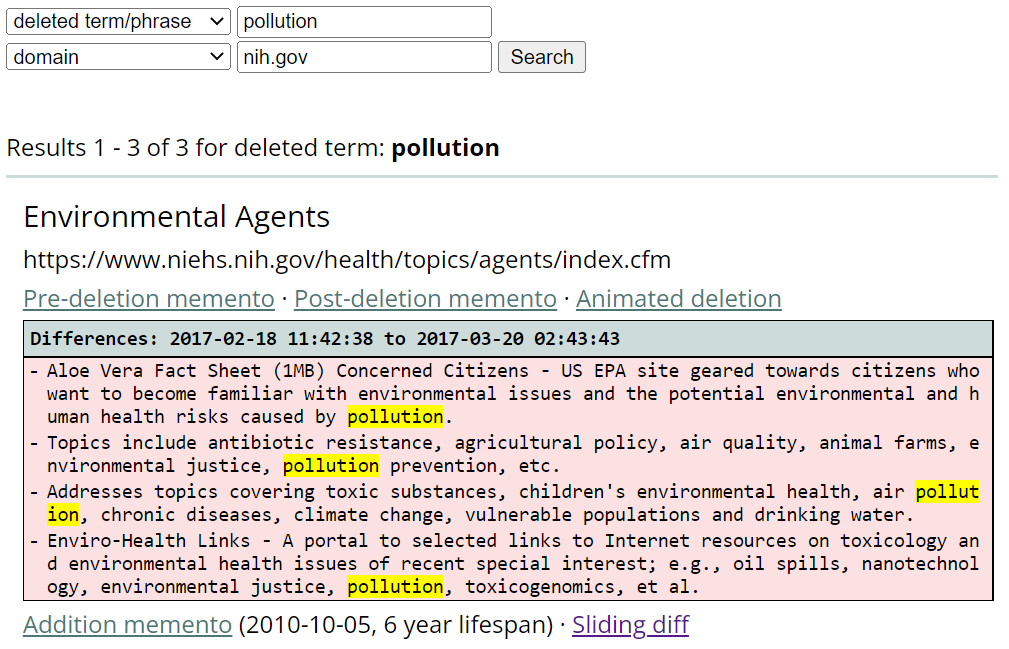}
         \caption{Change text SERP}
         \label{fig:vssolrnew}
     \end{subfigure}
        \caption{In SolrWayback, multiple versions of the same page are shown in the search results without any indication of how they are different. In the change text search engine results page, the versions of the page https://www.niehs.nih.gov/health/topics/agents/index.cfm that match the addition and deletion of the query term ``pollution" are indicated clearly.}
         \Description{In SolrWayback, two identical versions of the same page are shown consecutively in the search results page. It is not clear what differences exist between these pages. In fact, there are no content differences between the 2016 and 2017 versions, but there is some memento damage on the 2016 version. The version after the deletion of the search term is not returned in the results. In the change text search engine, the 2016 version is not shown. Instead, two versions from 2017 are shown, the version before the deletion and the version after.}
        \label{fig:vssolr2}
\end{figure*}

The Wayback Machine holds both paired mementos in the original EDGI set. In contrast, the new pair set has mementos located at multiple web archives. In this new set, there are 3,563 mementos with successful status codes in both 2016 and 2020. Table \ref{table:wa2016} shows the web archives that hold these mementos, according to their TimeMaps. Some URI-Rs have multiple valid mementos in the time range of interest, which is the first half of 2016 and 2020. In the table, these URI-Rs are counted once for each web archive. The percentages represent the number of URI-Rs with at least one memento found in the TimeMap in the specified time range, divided by 3,563. The percentages allow for comparison of holdings between archives and across years. Since each URI-R may have a memento at multiple archives, the sums of the columns are non-meaningful.

Firstly, the Library of Congress web archive holds mementos for many pages that are not available in the Wayback Machine in the first half of 2016. Future researchers examining U.S. federal websites should utilize the Library of Congress web archive in addition to the Wayback Machine. Secondly, it appears that Arquivo.pt conducted a crawl of U.S. federal websites in May 2016, and mementos with similar datetimes are not available in the Wayback Machine or at the Library of Congress. Using MemGator to query for aggregated listings of mementos across all web archives not only increases the amount of historical evidence available to analyze, but also brings light to entire collections and the possible motivations behind the creation of those collections. 

Indexing the paired 2016/2020 mementos is a starting point for determining additional versions of each page to index. The change text calculation script will determine all 
terms that have been added and removed between the two versions for each page indexed. Then, a binary search over the other versions of the page can be used to increase the temporal granularity of when a term was changed. Pages that were already identified as containing a term or phrase deletion by EDGI are prime candidates for early indexing. Since EDGI only tracked about 50 terms and phrases, additional meaningful terms can be identified from the change text calculations.

The initial indexing consisted of a small set consisting of 100 pairs of mementos. EDGI calculated that these pairs had complete or partial deletions of the terms sustainability, pollution, anthropogenic, or the phrase ``endangered species." The next indexing set was larger, consisting of 1,000 pairs, or 10\% of the original EDGI matched pairs. These pairs had complete or partial deletions of the terms and phrases: ``toxic", ``clean energy", ``climate change", and ``global warming". Many of the trends that emerged from the initial small sample persisted into the larger sample, so these trends are likely to persist throughout the entire data set. 

Downloading the 1,000 pairs using Hypercane, part of Level 1 on Figure \ref{fig:arch}, took 41 hours. Initial Lucene indexing of the WARCs using the UKWA WARC indexer took 2.5 hours, and PyWB indexing took 40 minutes. The change text calculation script took no more than 10 seconds to generate the JSON updates to post to the Lucene index, and posting the data took no more than 5 seconds. These steps correspond to Level 2 on Figure \ref{fig:arch}.

\subsection{Discussion}

The change text search engine results page has increased functionality compared to other temporal search engines for web archives, like SolrWayback, as well as compared to the web monitoring strategy used by EDGI. The change text search engine groups multiple versions of a page in a way that allows the user to make sense of the differences between them, as shown in Figure \ref{fig:vssolr2}. In SolrWayback, multiple identical versions of the same page will be shown in the search results page, and the version right after the deletion will not be a part of the search results. In contrast, the change text search engine results page only shows versions with meaningful change of the query term. The search engine results page can also be used to examine changes in context because of the diff snippet, and further detail is available in the sliding diff and animation tools linked in each result. Examining changes in context is not possible with the web monitoring strategy.

One of the major differences between indexing text content with Lucene and the deleted terms calculation completed by EDGI using Python is that the methods have completely different boilerplate removal techniques. Generally, the Lucene boilerplate removal strips more from the page than the EDGI technique. This meant that some terms were identified as deleted according to EDGI, but not according to the Lucene index. Other pages in Lucene indexed poorly due to the boilerplate removal, which is another difference in the deleted terms calculation. Some of the terms identified as deleted by EDGI were in various navigation page sections that were not stripped during boilerplate removal, so whether or not these should be named as deletions depends on the individual researcher's boilerplate removal preferences. Additionally, indexing all of the content with Lucene provides access to all of the deleted terms, while the EDGI web monitoring technique can only track pre-defined terms. The ability to find deleted terms without pre-defining them is powerful, which may outweigh poor indexing for some researchers.

While the original EDGI study only tracked 56 terms and phrases with environmental motivations, additional common deleted terms are now discoverable, as shown in Table \ref{table:newterms}. The original term list included terms like ``safety", ``transparency", ``regulation", and ``jobs", but ``public", ``access", ``action", ``development" and ``science" are terms with similar meaning that were also commonly deleted. Many of the top deleted terms were stop words or temporal terms. Two of the original EDGI terms, ``change" and ``state", are stop words that have meaning within a federal environmental dataset.\footnote{\href{https://countwordsfree.com/stopwords}{https://countwordsfree.com/stopwords}}

The list of the newly found terms, in order by most deletions, is: \texttt{national, support, public, program, resources, process, data, including, u.s, development, united, learn, \\ department, action, access, work, impact, tools, \\ areas, search, laboratory, technology, efforts, include, natural, science, planning, address,} and \texttt{open}.

\begin{table}[h!]
\centering
\begin{tabular}{ |c|c|c|  }
 \hline
 Term Type & Examples & Count \\
 \hline
 Stop words   & about, more, which, state    & 49 \\
 Temporal & 2015, 2, one, year & 13 \\
 EDGI & climate, clean, impacts, water & 11 \\
 Newly found & public, development, access, science & 29 \\
 \hline
 \end{tabular}
\caption{Categorization of top 100 deleted terms in the 1,000 paired memento index sample.}
\label{table:newterms}
\end{table}

Additional discovery of deleted phrases is possible by examining the search results for the new deleted terms. For example, ``public" is a new deleted term comparable to the original term ``transparency." Manual inspection of the search results for 
``public" showed that the term often comes up in the context of a deleted phrase. ``Public comment" was a common context for the deleted term ``public," and this usage is similar to the original term ``transparency." Another more frequent context was the phrase ``public health," which is actually more similar to the original term ``safety" than ``transparency." 

The two additional tools for examining changes in detail, the sliding diff and the animation, each have advantages for different contexts. The sliding diff tool can show more than two versions of a page, while the animation can only show two versions of a page at a time. The sliding diff also shows the changes in a persistent manner, while the deleted text in the animation view disappears by nature. Another benefit of the sliding diff tool is that it loads extremely quickly, because only content is compared. The animation must load both mementos through replay, compute the HTML difference, and then load and show the animation, so it takes much longer. The sliding diff shows the blocks of text with changes over time well, but diffs hide content that has not changed. If the user wants to view the entire page in context, the animation shows the entire page, including text and images that have not changed. The animation also makes the differences pop out more strongly than the sliding diff tool. Both tools suffer when there are too many changes on the page, because there is too much highlighted and the animation is too slow.

The EDGI federal webpages dataset was both valuable and appropriate for evaluating this change text search engine. Documents created by the government are not entitled to privacy; in fact, these documents should be archived and made discoverable for the people. However, researchers must take great care when indexing web archive collections. Heterogeneous web archive collections may contain material that should be subject to stronger privacy protections, such as websites made by minors. Lin et al. \cite{lin2020} speculated that people are unaware their websites are archived because they never gave explicit consent, and other people rely on a lack of a web archive search function to provide them with privacy. Mackinnon \cite{mackinnon2022databound} confirmed both of these hypotheses in a user study.
Both Lin et al. and Mackinnon suggest anonymization as a tool for researchers to study overall web archive collections without compromising the privacy of the content creators.

\section{Conclusions and Future Work}

Existing temporal search engines for web archives do not allow for users to query for change over time. This paper presents a change text search engine, which allows users to find and view the changes in webpages. The search engine results page groups multiple versions of a page together without hiding the changes between these versions. In fact, the changes between the versions are the core reason why the grouping can occur, since the grouping represents the lifespan of a term on the page. A deletion animation shows changes in context, and a sliding difference viewer enables quick examination of the differences between many versions a page. The inverted index contains valuable information about the most frequently deleted terms in the corpus.

One important aspect of future work regarding the architecture will include automating the indexing process. Kiesel et al. \cite{Kiesel2018} automated their personal web archiving indexing for both Lucene and PyWB. The change text calculation script will also need to be automated for this system to function. Automating the dual Lucene/PyWB indexing was identified as a significant architecture advancement for the full-text search in web archives community by Jackson et al. \cite{jackson2022iipc} Another improvement to the search engine code will involve expanded support for detecting added phrases when both terms are already present on the page separately, and identification of partially added terms. 

Indexing additional earlier versions of the webpages will lead to important insights about when terms were added, such 
as during a
prior Presidential administration. Earlier versions will also enable discovery of additional terms and phrases that were deleted in prior administrations. Viewing the index from a computational standpoint allowed for the calculation of new deleted terms. Additional work will include computation of new deleted phrases. Another aspect of future work will include categorizing the amount of change to a webpage. This work would make it possible to distinguish between different types of changes, such as an entire page rewrite aligned with an organization's new goals versus removals of blocks of content that remove public access to vital information.

The need for a change text search engine and the specific search engine result page features was justified by examining specific user tasks. Jayanetti et al. \cite{jayanetti-tpdl22a} categorized users in web archives, including those who view multiple versions of the same page. Future work on the user interface will involve analyzing the tasks of these users to better determine what these users need. Future work will also include a user study of the search interface front end, including the animation, to evaluate whether users such as journalists can more successfully complete their information seeking tasks.



\bibliographystyle{ACM-Reference-Format}
\bibliography{myref}


\end{document}